\documentclass[11pt]{article}
\pdfoutput=1
\usepackage{graphicx}	
\usepackage{amsmath}	
\usepackage{dcolumn}
\usepackage{bm}
\usepackage{graphics}
\usepackage{afterpage}
\usepackage{float}
\usepackage{subfigure}
\usepackage{rotating}
\usepackage{multirow}
\usepackage{tabularx}
\usepackage{booktabs}
\usepackage{multirow}
\usepackage{fancyhdr}
\usepackage[utf8]{inputenc}
\usepackage{theorem}
\usepackage{moreverb}
\usepackage{euscript}
\usepackage{psfrag}
\usepackage{slashed}
\usepackage{mathtools}
\usepackage{makecell}
\usepackage[flushleft]{threeparttable}
\usepackage{adjustbox}

\usepackage{dcolumn}
\usepackage{bm}
\usepackage[mathlines]{lineno}
\usepackage{lettrine}
\usepackage[dvipsnames]{xcolor}
\usepackage{jhep}
\usepackage{listings}
\usepackage{color,xcolor}
\usepackage{natbib}

\input Zallman.fd

\LettrineTextFont{\itshape}
\usepackage{hyperref}
\hypersetup{
     colorlinks   = true,
     citecolor    = Cyan,
     urlcolor     = Cyan,
     linkcolor    = Cyan
}

\preprint{SLAC-PUB-17741}

\title{Milky Way White Dwarfs as Sub-GeV to Multi-TeV Dark Matter Detectors}

\author[a]{Javier F. Acevedo,}
\emailAdd{jfacev@slac.stanford.edu}
\author[a,b]{Rebecca K. Leane,}
\emailAdd{rleane@slac.stanford.edu}
\author[a]{Lillian Santos-Olmsted,}
\emailAdd{solmsted@stanford.edu}

\affiliation[a]{Particle Theory Group, SLAC National Accelerator Laboratory, Stanford, CA 94035, USA}
\affiliation[b]{Kavli Institute for Particle Astrophysics and Cosmology, Stanford University, Stanford, CA 94035, USA}

\date{\today}
\abstract{We show that Milky Way white dwarfs are excellent targets for dark matter (DM) detection. Using Fermi and H.E.S.S. Galactic center gamma-ray data, we investigate sensitivity to DM annihilating within white dwarfs into long-lived or boosted mediators and producing detectable gamma rays. Depending on the Galactic DM distribution, we set new constraints on the spin-independent scattering cross section down to $10^{-45}-10^{-41}$~cm$^2$ in the sub-GeV DM mass range, which is multiple orders of magnitude stronger than existing limits. For a generalized NFW DM profile, we find that our white dwarf constraints exceed spin-independent direct detection limits across most of the sub-GeV to multi-TeV DM mass range, achieving sensitivities as low as about $10^{-46}$~cm$^2$. In addition, we improve earlier versions of the DM capture calculation in white dwarfs, by including the low-temperature distribution of nuclei when the white dwarf approaches crystallization. This yields smaller capture rates than previously calculated by a factor of a few up to two orders of magnitude, depending on white dwarf size and the astrophysical system.}

\begin{document}
\maketitle

\newpage
\section{Introduction}

\lettrine{O}{ur milky way galaxy} is host to hundreds of billions of stars. Unfortunately, these stars can not shine brightly forever, and they must eventually meet their demise. The final evolutionary state for over 97 percent of stars is called a \textit{white dwarf}. As a star enters its red giant phase, it fuses helium to carbon and oxygen in its core, but if it does not have sufficient mass to generate the temperatures required for carbon fusion, an inert mass of carbon and oxygen will accumulate at its center. White dwarfs are the remnants of low to medium mass stars that have exhausted their nuclear fuel, and shed their outer layers, revealing the remaining core, typically composed of a crystalline lattice of carbon and oxygen atoms. 

White dwarfs are incredibly dense, packing the mass of our Sun into a sphere the size of the Earth. It is exactly these two properties -- density and radius -- which make it an ideal target for new dark matter (DM) searches. Its density renders it one of the most dense celestial objects in the Universe, only coming second to neutron stars. But compared to neutron stars, white dwarfs are much larger, such that the surface area available for DM capture is much larger. This aids DM detection, as larger maximal capture rates mean more potential consequent particles arising from DM annihilation in white dwarfs. Furthermore, the sheer number of white dwarfs present in our Milky Way Galaxy is about one order of magnitude higher than the number of neutron stars, due to the prevalence of low to medium mass stars and their final evolutionary fate.

White dwarfs have been considered as DM detectors in the past~\cite{Mochkovitch:1985vi, Moskalenko:2006mk,Bertone:2007ae,McCullough:2010ai,Hooper:2010es,
Kouvaris:2010jy,
Bramante:2015cua,
Graham:2015apa,
Amaro-Seoane:2015uny,
Graham:2018efk,
Cermeno:2018qgu,
Acevedo:2019gre,
Dasgupta:2019juq,
Acevedo:2020avd,
Acevedo:2021kly,
Janish:2019nkk,
Krall:2017xij,
Panotopoulos:2020kuo,
Curtin:2020tkm,
Bell:2021fye,
DeRocco:2022rze,
Ramirez-Quezada:2022uou,
Smirnov:2022zip,
Garani:2023esk}. The most commonly considered scenario for sub-GeV DM, is DM capture and consequent annihilation to short-lived or not very boosted mediators, which in turn leads to the DM rest mass energy being absorbed by the white dwarf. In this case, the temperature of the white dwarf can increase proportionally to the density of DM available in its local environment. As white dwarfs are generally quite hot, a large local DM density is required to achieve an appreciable DM heating signal. Therefore, previous work has dominantly focused on white dwarf DM heating in globular clusters such as Messier 4 (M4)~\cite{Bertone:2007ae,McCullough:2010ai,Amaro-Seoane:2015uny,
Cermeno:2018qgu,
Krall:2017xij,Dasgupta:2019juq,
Panotopoulos:2020kuo,Bell:2021fye,DeRocco:2022rze,Ramirez-Quezada:2022uou,Garani:2023esk}, where sufficiently cool white dwarfs have been observed, and the DM density has been speculated to potentially be as high as about $\mathcal{O}$(100) GeV/cm$^3$~\cite{Bertone:2007ae,McCullough:2010ai}; about 1,000 times higher than the local Solar position. However, as was already emphasized in the bulk of these studies, the DM density in globular clusters such as M4 is not well understood, and the current systematic error bar on the DM density is substantial, with some astrophysical studies long stating consistency with little to no DM at all~\cite{Moore:1995pb,Saitoh:2005tt}. Therefore, while the commonly-studied white dwarf heating signal offers substantial potential sensitivity, it is not yet a robust search.

In this work, we investigate a new search for DM in white dwarfs. We target white dwarfs in the inner region of the Milky Way Galaxy, where they are numerous, and where the DM content is still high and yet much better understood ($i.e.$, the DM density is not expected to be zero in the Galactic center). While the inner slope of the DM distribution in our Galaxy is not yet known, we consider a range of variations to somewhat bracket the expected systematics, which still give substantial amounts of DM to power a signal in white dwarfs. The scenario we investigate is also complementary to DM heating searches, which require mediators to not escape the white dwarf; we investigate the scenario where mediators decay outside the white dwarf and produce gamma rays. As an important component of this work, we also present and use an improved calculation for DM capture in white dwarfs, which takes into account the velocity distribution of low-temperature nuclei in the white dwarf as it approaches crystallization. Depending on the white dwarf size and the astrophysical system of interest, this produces white dwarf capture rates between a factor of a few up to about two orders of magnitude smaller than previously calculated.

This paper is organized as follows. In Section~\ref{sec:wd_modeling}, we discuss the white dwarf properties relevant for our capture calculation, including our modeling of the Galactic white dwarf radial number distribution, as well as their physical properties such as temperature and composition. In Section~\ref{sec:capture}, we discuss our improved treatment for DM capture in white dwarfs, building on the existing literature. In Section~\ref{sec:detection}, we discuss the detection prospects for DM annihilation to gamma rays in these white dwarfs with Fermi and H.E.S.S.. We show our new limits on the spin-independent DM-SM scattering cross section in Section~\ref{sec:DMconstraints}, and include some additional discussion on the relevant astrophysical uncertainties for our search. We summarize our findings and conclude in Section~\ref{sec:conclusion}. Appendix~\ref{app:RMSvel} contains additional discussion on the low-temperature ion velocity in white dwarfs.

\section{White Dwarf Properties and Galactic Distribution}
\label{sec:wd_modeling}

\subsection{Distribution in the Inner Milky Way}
\label{sec:wd_dist}
To determine the size of our white dwarf population DM signal, we first need to know how many white dwarfs reside in the inner Galaxy, as well as how they are distributed. Our signal is dominated by the central region of the Galaxy, as it is where the density of white dwarfs and the density of DM are at their highest. Stars in the inner parsec of our Galaxy are a part of what is called the nuclear star cluster, which surrounds the central supermassive black hole of the Milky Way, Sagittarius A$^*$ (Sgr~A$^*$), and contains a significant stellar density. 

We parameterize the white dwarf number density distribution $n_{\rm WD}$ in this central region as a power-law with slope $\alpha$,
\begin{equation}
    n_{\rm WD}(r) = n_{\rm WD}(r_0) \left(\frac{r}{r_0}\right)^{-\alpha}~,
    \label{eq:nsc_pw}
\end{equation}
where $r$ is the radial distance to the Galactic center, and $r_0$ and $n_{\rm WD}(r_0)$ define the normalization as we will discuss shortly. Power-law density profiles for the stellar components of nuclear star clusters are predicted by $N$-body simulations~\cite{Amaro-Seoane:2010dzj,2018A&A...609A..28B,Panamarev:2018bwq,Zhang:2023cip}, as well as analytical mass segregation solutions assuming a steady state is reached through gravitational interactions between the stars \cite{1977ApJ...216..883B,Hopman_2006,Alexander:2008tq}. Cuspy distributions of the stellar objects in the inner parsec surrounding Sgr~A$^*$ are also favoured by recent spectroscopic and photometric surveys \cite{2018A&A...609A..26G,2018A&A...609A..27S,2019ApJ...872L..15H} and X-ray observations \cite{2018Natur.556...70H}. The index depends on several modeling assumptions including the star formation history, initial mass function, and initial-to-final mass relation for stellar remnants. The latter in particular can be sensitive to the exact metallicity value; we detail this further below. These assumptions determine the distribution of masses and abundance of each star type, which in turn influences how the energy is distributed among the components of the stellar population through gravitational scattering, and therefore their number density at a given radius. We adopt $\alpha = 1.4$ following the analytic estimates of Ref.~\cite{Alexander:2008tq}.

To determine the normalization of Eq.~\eqref{eq:nsc_pw}, we use the state-of-the-art predictions for the number of white dwarfs as per Ref.~\cite{2023ApJ...944...79C}. There, a number of star formation scenarios in the Galactic center were analyzed, leading to a prediction of $\sim 8.7 \times 10^6$ white dwarfs within the central $1.5 \ \rm pc$. Thus, with the exponent choice above, fixing $r_0 = 1.5 \ \rm pc$ gives $n_{\rm WD}(r_0) \simeq 3.28 \times 10^5 \ \rm pc^{-3}$. This normalization assumes the power-law is only valid between $10^{-4} \ \rm pc$ and $1.5 \ \rm pc$, which matches the approximate range of Ref.~\cite{Alexander:2008tq}.
This estimate is a factor of $\sim 4-10$ larger than previous white dwarf counts in the Galactic center, see $e.g.$ Refs.~\cite{Hopman_2006,Alexander:2008tq}.
This is because Ref.~\cite{2023ApJ...944...79C} incorporated for the first time metallicity measurements from spectroscopic surveys \cite{2015ApJ...809..143D,2015A&A...573A..14R,2017MNRAS.464..194F}, which revealed a large population of stars with super-solar metallicity in this region. Higher metallicity leads to enhanced mass loss rates during the progenitor phase through stellar winds, and as a result more of these compact objects are produced compared to lower metallicity regions. 

\subsection{Age, Temperature, and Composition}
\label{sec:wdcomp}

As well as knowing how the white dwarf population is distributed, we need to consider the range of properties expected for the individual white dwarfs. We model the dense white dwarf plasma using the Feynman-Metropolis-Teller equation of state \cite{Feynman:1949zz},  which in its most recent iteration incorporates a number of corrections ranging from weak and nuclear interactions to general relativistic effects \cite{Rotondo:2011zz,2011PhRvC..83d5805R}. This equation of state is commonly used in previous works on DM capture in white dwarfs, and therefore also facilitates direct comparison.
For a given central density and composition, the resulting white dwarf configuration is obtained by coupling the equation of state to the Tolman-Volkoff-Oppenheimer equation (see $e.g.$ Ref.~\cite{Shapiro:1983du}).

The exact composition of white dwarfs can depend on a number of its progenitor's properties, including its mass, rotation, metallicity, and binarity. Both observations and simulations of the white dwarf mass distribution indicate that most of these objects have masses $0.3 \, M_{\odot} \lesssim M_{\rm WD} \lesssim 1.0\, M_\odot$ (where $M_\odot$ is the Solar mass), for which they are expected to be predominantly made of carbon-12 and oxygen-16 \cite{2012ApJS..199...29G,2016MNRAS.461.2100T}. For white dwarfs with masses $M_{\rm WD} \lesssim 0.3 \, M_{\odot}$, their composition is predominantly helium-4. However, it is well known these low-mass remnants cannot be the result of single star evolution and are thus rare, see $e.g.$ Ref.~\cite{2019ApJ...871..148L}. In the case of ultra-massive white dwarfs with $M_{\rm WD} \gtrsim 1.0 \, M_{\odot}$, a fraction of this population will also be composed of carbon-12 and oxygen-16 as a result of mergers of binary white dwarfs \cite{Yoon:2007pw,2009AIPC.1122..320L,Raskin:2011aa} or the specific evolution of its progenitor \cite{1996ApJ...472..783D,2019NatAs...3..408D,2021A&A...646A..30A}, with the remaining white dwarfs having oxygen-16 cores with traces of neon-20 and heavier elements. 
In fact, $N$-body simulations predict carbon-oxygen white dwarfs to be more prevalent in our central Galactic region of interest~\cite{Panamarev:2018bwq}.
To be conservative, we assume our white dwarfs are only made of carbon-12. This is conservative because, while the equation of state used predicts extremely similar density profiles and mass-radius relations for both pure carbon and pure oxygen compositions \cite{Rotondo:2011zz}, DM-oxygen scattering has a larger coherent enhancement factor compared to carbon, which would produce larger capture rates.
Furthermore, we emphasize that the equation of state considered here predicts very similar profiles to other existing equations of state such as Ref.~\cite{1961ApJ...134..683H}, except near the stability threshold where it actually predicts smaller masses compared to other treatments. Thus, the limits we derive from the white dwarfs in the high mass end are also conservative. 

Table~\ref{tab:wds} details the three benchmark white dwarfs we consider: one light mass, one intermediate mass, and one heavy mass white dwarf. These are chosen to span the range of realizations of white dwarfs expected in galaxies and globular clusters, and to facilitate comparison with the results in Ref.~\cite{Bell:2021fye}. We also assume that the white dwarfs powering the gamma-ray signal from DM annihilation have ages greater than about 3 Gyr. Recent star formation history analyses of the Milky Way's nuclear star cluster show preference for a double-starburst model, with the initial star formation episode occurring between $5 - 13 \ \rm Gyr$ ago and producing over $\sim 90 \%$ of the stars observed today \cite{2020A&A...641A.102S,2023ApJ...944...79C}. Thus, the majority of white dwarfs around or above a solar mass must have formed by now, since their progenitors only last less than about 100 Myr. For our light mass benchmark, on the other hand, their existence is contingent upon the precise time at which the first starburst occurred, since these white dwarfs are formed from roughly solar mass progenitors which can last for about 10 Gyr. If an early starburst occurred $\sim 13 \ \rm Gyr$ ago \cite{2020A&A...641A.102S}, then these stars must have ages greater than about 3 Gyr. The most advanced white dwarf cooling models predict white dwarfs in this age range will have a luminosity of order $\lesssim 10^{-4} \, L_{\odot}$ \cite{2022MNRAS.509.5197S}, where $L_\odot$ is the solar luminosity. This corresponds to white dwarf core temperatures $T_{\rm WD} \lesssim 10^6 \ \rm K$, with some slight variation depending on the atmospheric composition \cite{Shapiro:1983du}.

\begin{table}
\centering{
\renewcommand{\arraystretch}{1.5}
\begin{tabular}{|c|c|c|c|} 
 \hline
 \textsc{Benchmark} & \textsc{Mass [$M_\odot$]}  & \textsc{Radius [$\rm km$]} & \textsc{Central Density [$\rm g / cm^{3}$]} \\ [0.5ex] 
 \hline
  Light Mass & $0.49$ & $9390$ & $1.98 \times 10^6$\\ 
 \hline
  Intermediate Mass & $1.00$ & $5380$ & $3.46 \times 10^7$\\ 
 \hline
  Heavy Mass & $1.38$ & $1250$ & $1.14 \times 10^{10}$\\ 
 \hline
\end{tabular}}
\caption{Properties of our benchmark white dwarfs. We assume all the benchmarks to be pure carbon-12; see text for details.}
\label{tab:wds}
\end{table}

\section{Refined Treatment of Dark Matter Capture in White Dwarfs}
\label{sec:capture}

Equipped with the properties and number distribution of white dwarfs in the inner Galaxy, we now discuss the rates in which white dwarfs capture DM. We will focus on DM capture through scattering against white dwarf ions, and make some improvements on the previous treatments of DM capture in white dwarfs. 

\subsection{Velocity Distributions}
\label{sec:veldist}
We first consider the distribution of the relative velocity between the DM and the target ions, which is an important quantity for DM capture. This is given by  
\begin{equation}
    \mathcal{F}_{\rm rel}(u_{\chi}) = \frac{u_\chi}{v_{\rm WD}} \left(\frac{3}{2\pi (v_d^2 + \langle v_N^2 \rangle)}\right)^{1/2} \left[\exp\left(-\frac{3(u_\chi-v_{\rm WD})^2}{2 (v_d^2 + \langle v_N^2 \rangle)}\right)-\exp\left(-\frac{3(u_\chi+v_{\rm WD})^2}{2 (v_d^2 + \langle v_N^2 \rangle)}\right)\right]~,
    \label{eq:relvdist}
\end{equation}
where $u_\chi$ is the DM-ion relative velocity before the DM is accelerated by the white dwarf, $v_d$ is the local velocity dispersion of the DM halo, $v_{\rm WD}$ is the white dwarf speed in the Galactic rest frame, and $\langle v_N^2 \rangle$ is the mean squared white dwarf ion velocity. We assume a constant DM velocity dispersion throughout the Galaxy equal to $v_d = v_d^{\odot} \simeq 270 \ \rm km / s$; the dispersion towards the Galactic center is not precisely known, and in the inner region ($10^{-2}-1$~pc) approximately varies in the range $\sim 200 - 900 \ \rm km/s$ \cite{2013PASJ...65..118S}. Even taking this large range of DM velocities translates only into a factor of about two in variation in the capture rates we derive. For the white dwarf speed $v_{\rm WD}$, we adopt a value of $v_{\rm WD} = \sqrt{8/\pi} \, \sigma_* \simeq 285 \ \rm km / s$, where $\sigma_* = 179 \rm \ km / s$ is the inferred velocity dispersion for old stars in the Milky Way's nuclear stellar cluster \cite{Trippe:2008vj,2009A&A...502...91S}. In practice, the specific choice of these parameters does not have a significant impact on our results due to the magnitude of $\langle v_N^2 \rangle$, which is the dominant scale in the above distribution for most white dwarfs in the mass range we analyze. Note that Eq.~\eqref{eq:relvdist} does not explicitly include a velocity cutoff at the Galactic escape velocity, which is expected to be of order $\sim 1000 \ \rm km/s$ in the central inner parsec \cite{2013A&A...549A.137I}. We have checked that our capture rates are fairly insensitive to the inclusion of such cutoff, changing only by a few percent in the regime where most of the distribution above is relevant for capture. 

To estimate the mean squared ion velocity $\langle v_N^2 \rangle$, we must first discuss the velocity distribution that results from the microscopic structure of a white dwarf. Most of a white dwarf volume is fully ionized due to the extreme pressure, with the state of such a dense plasma depending on the ratio between the Coulomb energy of the ions and the temperature. When the white dwarf is formed, the ions are hot enough to form a dense Boltzmann gas. This phase smoothly transitions into a strongly coupled liquid as the object cools down. This liquid plasma phase eventually solidifies into a lattice, in a first-order phase transition known as white dwarf crystallization. The transition point for a single component plasma was first numerically estimated in Ref.~\cite{1975ApJ...200..306L}. More detailed and accurate numerical calculations (see $e.g.$ Ref.~\cite{2021PhRvE.103d3204B}) are in reasonable agreement with this estimate, and empirical confirmation of this first-order phase transition has been obtained from Gaia data in Ref.~\cite{2019Natur.565..202T}.

Even prior to the onset of crystallization, at sufficiently low temperatures each ion feels a restoring force from the neighboring ions which can be approximated as a harmonic oscillator potential, with a natural constant given by the ion plasma frequency,
\begin{equation}
    \omega_p = \left(\frac{4 \pi Z^2 e^2 n_N}{m_N}\right)^{1/2} \simeq 1.68 {\ \rm keV} \left(\frac{\rho_{\rm WD}}{1.98 \times 10^6 \ \rm g / cm^{3}}\right)^{1/2} \left(\frac{Z}{6}\right) \left(\frac{A}{12}\right)^{-1}\,,
    \label{eq:ion_plasma_freq}
\end{equation}
where $n_N$ is the white dwarf ion number density, $m_N = A \, m_n$ is the ion mass with $m_n = 0.93 \ \rm GeV$, $A$ is the atomic mass, and $Z$ is the atomic number. On the right of Eq.~(\ref{eq:ion_plasma_freq}) we have normalized to the central value of our light benchmark white dwarf. This approximation will be valid for temperatures that satisfy $ T_{\rm WD} \lesssim \omega_p$. In this regime, ion quantum effects also become relevant, as the thermal de Broglie wavelength of the ions is much shorter than their average separation. As we show in Appendix~\ref{app:RMSvel}, the velocity distribution of the ions in this regime is well approximated by
\begin{figure*}[t!]
    \centering
    \includegraphics[width=\textwidth]{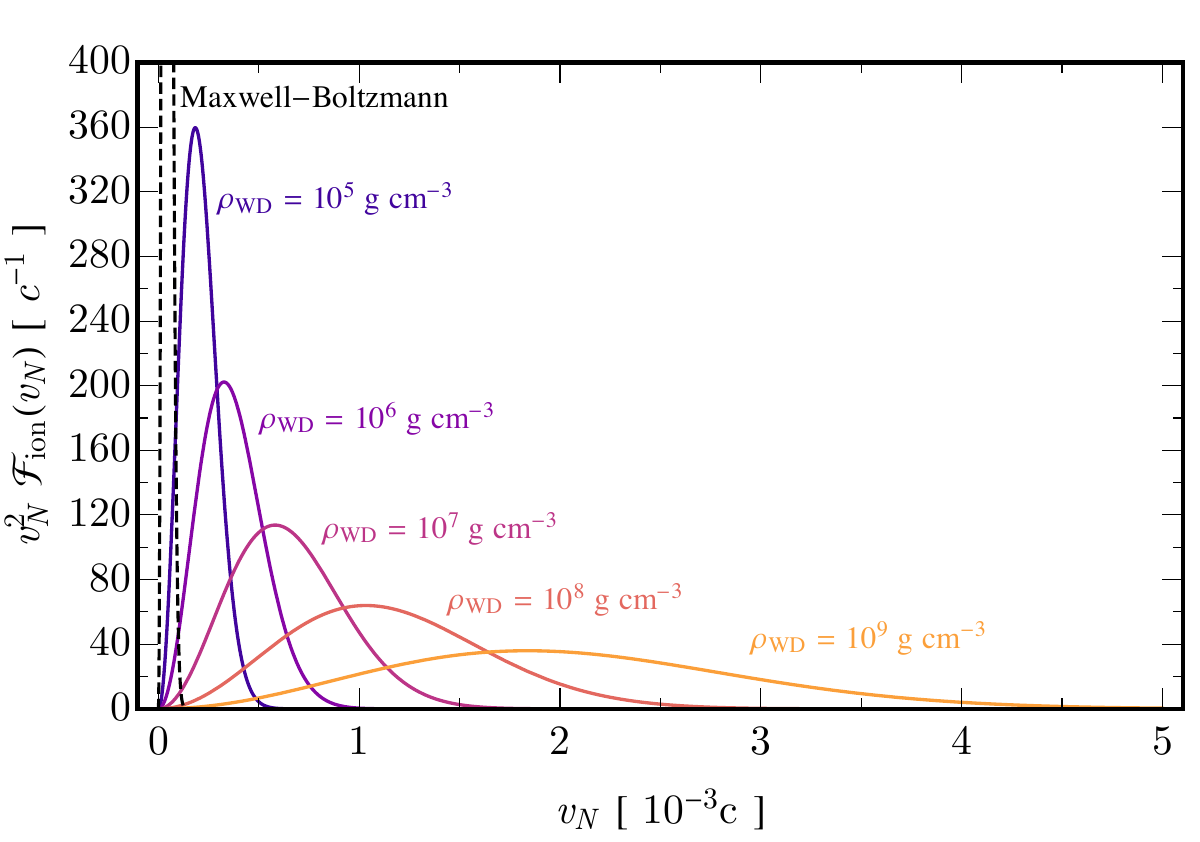}
    \caption{Speed probability density of the white dwarf ions in the stellar rest frame using the classical Maxwell-Boltzmann distribution (\textbf{dashed}), and the more accurate distribution accounting for quantum effects which produce higher ion velocities at the specified densities (\textbf{solid}). For this plot, we assume a temperature $T_{\rm WD} = 10^5 \ \rm K$ and a pure carbon-12 composition.}
    \label{fig:momdist}
\end{figure*}

\begin{equation}
     \mathcal{F}_{\rm ion}(v_N)= \left(\frac{m_N}{\pi \omega_p \coth{\left(\omega_p /2 T_{\rm WD}\right)}}\right)^{3/2} \exp\left[-\frac{m_N v_N^2}{\omega_p \coth\left(\omega_p/2T_{\rm WD}\right)}\right]~,
     \label{eq:momdist-main}
\end{equation}
which gives a mean squared velocity
\begin{equation}
    \langle v_N^2 \rangle = \left(\frac{3\omega_p}{2m_N}\right) \coth \left(\frac{\omega_p}{2T_{\rm WD}}\right)~.
    \label{eq:veldisp-main}
\end{equation}
In what follows, we will assume that white dwarfs have a uniform temperature $T_{\rm WD}$. This is a good approximation for the volume relevant for capture in the optically thin limit, where degenerate electrons have high thermal conductivity.

Figure~\ref{fig:momdist} shows our velocity distribution for a  temperature $T_{\rm WD} = 10^5 \ \rm K$ and density values, which translates into ion plasma frequencies of order $\sim 0.37 - 40 \ \rm keV$. These are chosen to approximately cover the range of our white dwarf benchmarks ($cf.$~Table~\ref{tab:wds}). For comparison, we show as the dashed line the classical Maxwell-Boltzmann distribution at the same temperature of $T_{\rm WD} = 10^5 \ \rm K$. As the density decreases, as is expected towards the white dwarf surface, eventually our distribution converges to the Maxwell-Boltzmann because quantum effects stop being relevant at low ion plasma frequencies. It can therefore be seen that our approach naturally reproduces the velocity distribution used in Ref.~\cite{Bell:2021fye}, if we consider the outer regions of the object where $T_{\rm WD} \gg \omega_p$. In this limit, $\langle v_N^2 \rangle$ converges to the expected limit from classical statistical mechanics, $i.e.$ $3T_{\rm WD}/m_N$. We further expand on this point in Appendix~\ref{app:RMSvel}. In practice, however, DM capture in the optically thin limit will occur within the volume where the mean squared velocity scales as $\sim 3\omega_p/2m_N$, which is significantly greater than the naive thermal value. As we discuss further below, this discrepancy can produce significant deviations compared to the previous treatment. This ranges from an order one factor in regions where the DM has a high velocity dispersion, up to about two orders of magnitude in systems with low velocity dispersion such as globular clusters where $v_d^2 \ll \langle v^2_N \rangle$.

\subsection{Capture Rate for a Single White Dwarf}

\subsubsection{Formalism}

We now describe the DM capture rate in a single white dwarf. We follow the treatment of DM in white dwarfs considered in Ref.~\cite{Bell:2021fye}, which is based on the earlier Refs.~\cite{1987ApJ...321..571G,Gould:1987ju,Busoni:2017mhe} (which had been applied to the Earth and Sun), however we improve upon the implementation of the velocity distribution of the white dwarf ions, accounting for quantum effects at low temperatures as discussed in the previous section. In the optically thin limit, where the DM particles scatter on average less than once per white dwarf crossing, the capture rate $C_{\rm WD}$ is given by
\begin{equation}
    C_{\rm WD} = \frac{\rho_{\chi}}{m_\chi} \int_0^{R_{\rm WD}} dr \, 4 \pi r^2 \int_0^{\infty} du_\chi \, \frac{w(r)}{u_{\chi}} \mathcal{F}_{\rm rel}(u_{\chi}) \, \Omega^{-}(w)\, ,
    \label{eq:cap1}
\end{equation}
where the down-scattering rate is
\begin{equation}
    \Omega^{-}(w) = \int_0^{v_{\rm esc}(r)} dv \, R^{-}(w \rightarrow v)\, ,
    \label{eq:cap2}
\end{equation}
and the differential down-scattering rate is
\begin{equation}
    R^{-}(w \rightarrow v) = \int_0^{\infty} ds \int_0^{\infty} dt \, {32 \pi \mu_{+}^4} \frac{v t}{w} \frac{d\sigma_{N\chi}}{d\cos\theta} \, n_N(r) \mathcal{F}_{\rm ion}(v_N) \Theta\left(t+s-w\right) \Theta\left(v-|t-s|\right)\, ,
    \label{eq:cap3}
\end{equation}
with the following definitions
\begin{equation}
    w(r)^2 = v_{\rm esc}(r)^2+u_\chi^2~,
    \label{eq:omegar}
\end{equation}
\begin{equation}
    v_N(r)^2 = 2\mu\mu_+ t^2 + 2\mu_+ s^2 - \mu w(r)^2~,
\end{equation}
\begin{equation}
    \mu = \frac{m_\chi}{m_N}~,
\end{equation}
\begin{equation}
    \mu_{\pm} = \frac{\mu \pm 1}{2}~,
\end{equation}
and the distributions $\mathcal{F}_{\rm rel}(u_\chi)$ and $ \mathcal{F}_{\rm ion}(v_N)$ are given by Eqs.~\eqref{eq:relvdist} and \eqref{eq:momdist-main} respectively. Note that our expression also differs from that of Ref.~\cite{Bell:2021fye} by a factor $\pi^{3/2}$ due to the way we have normalized $ \mathcal{F}_{\rm ion}(v_N)$. The factors $m_\chi$ and $\rho_\chi$ respectively are the DM mass and local DM mass density around the white dwarf. The total DM particle capture rate $C_{\rm WD}$ is obtained through the integration over shells of the rate at which DM particles scatter and become gravitationally bound to the white dwarf. The velocities $w$ and $v$ denote the DM velocity in the white dwarf rest frame before and after scattering, respectively. For capture, we require $v \lesssim v_{\rm esc}(r)$ at the specific shell where the scattering occurred. Lastly, the variables $s$ and $t$ are the center-of-mass velocity and the velocity of the DM in the center-of-mass frame, respectively.
  
We consider an isotropic and constant DM-ion cross section, given by
\begin{equation}
    \frac{d\sigma_{N\chi}}{d\cos\theta} = A^4 \left(\frac{m_\chi+m_n}{m_{\chi}+A m_n}\right)^2 |F_{\rm helm}(E_R)|^2 \, \sigma_{n\chi}\,,
\end{equation}
where $\sigma_{n\chi}$ is the DM-nucleon cross section. The function $F_{\rm helm}(E_R)$ is the Helm form factor, which accounts for nuclear substructure effects for scattering at large momentum transfers compared to the inverse nuclear radius. This function can be expressed in terms of the recoil energy of the target
\begin{equation}
    |F_{\rm helm}(E_R)|^2 = \exp\left(-\frac{E_R}{E_N}\right)~,
\end{equation}
where $E_R = q^2/2m_N$ is the recoil energy, and $q$ is the magnitude of the momentum transfer. In terms of the center-of-mass velocity $s$ and the velocity of the DM $t$ in this frame, the average momentum transfer magnitude reads \cite{Busoni:2017mhe}
\begin{equation}
    q(s,t) = \left[2m_\chi^2 t^2 \left(1-\frac{(s^2+t^2-v^2)(s^2+t^2-w^2)}{4s^2t^2}\right)\right]^{1/2}\, .
\end{equation}
The scale $E_N = 3/(2m_N\Lambda_N^2)$ determines the recoil energy at which coherence loss becomes relevant, where we parameterize $\Lambda_N$ as \cite{Helm:1956zz}
\begin{equation}
    \Lambda_N \simeq \left(0.91 \left(\frac{m_N}{\rm GeV}\right)^{1/3}+0.3\right) \, \rm fm~.
\end{equation}
We integrate the capture rate $C_{\rm WD}$, as given by Eq.~\eqref{eq:cap1}, using the numerical code \texttt{Vegas}~\cite{PETERLEPAGE1978192}.

The capture rate given by Eq.~\eqref{eq:cap1} is directly proportional to the background DM density $\rho_{\chi}$, which is of course dependent on the position within the Galaxy. We consider the following DM halo profiles
\begin{equation}
    \rho_{\chi}^{\rm (Ein)}(r) = \rho_{\chi}^0 \exp\left[-\frac{2}{\alpha} \left( \left(\frac{r}{r_s}\right)^{\alpha} - 1\right)\right]~,
\end{equation}

\begin{equation}
    \rho_{\chi}^{\rm (NFW)}(r)  = \dfrac{\rho_{\chi}^0}{\left(\dfrac{r}{r_s}\right)^{\gamma} \left(1+\dfrac{r}{r_s}\right)^{3-\gamma}}~,
\end{equation}
where $\rho_{\chi}^{\rm (Ein)}(r)$ is the Einasto halo profile, and $\rho_{\chi}^{\rm (NFW)}(r)$ is the Navarro-Frenk-White (NFW) halo profile. We will consider the Einasto profile as the core-like DM profile benchmark with $\alpha=0.17$ and $r_s=20$~kpc, which matches well for Milky Way-like halos in the DM-only Aquarius simulations~\cite{Pieri:2009je}. We will also study an NFW profile which fixes $\gamma=1$, as well as a more cuspy generalized NFW (gNFW) profile with $\gamma=1.5$, motivated by the expected adiabatic contraction in the inner Galaxy~\cite{2011arXiv1108.5736G,DiCintio:2014xia}. For both of the NFW-like profiles we set $r_s=12$~kpc. For all DM density profiles, the normalization constant $\rho_\chi^0$ is adjusted so that the profile evaluates to $\rho_\chi^\odot \simeq 0.42 \ \rm GeV / cm^{3}$ in the Solar neighborhood.

\subsubsection{Comparison with Previous Treatment of the Capture Rate}

\begin{figure}[t!]
    \centering
    \includegraphics[width=0.495\textwidth]{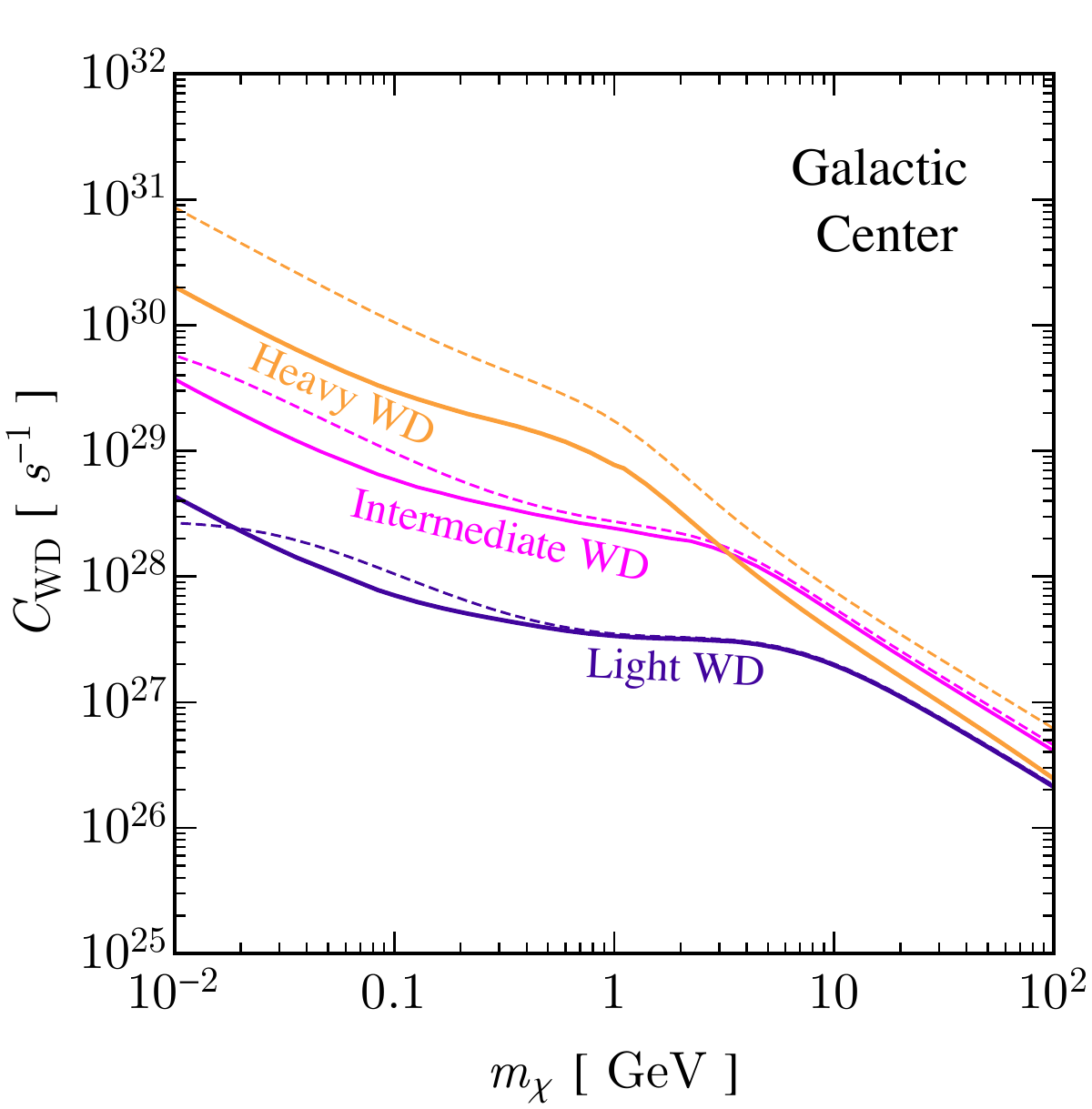}
    \includegraphics[width=0.495\textwidth]{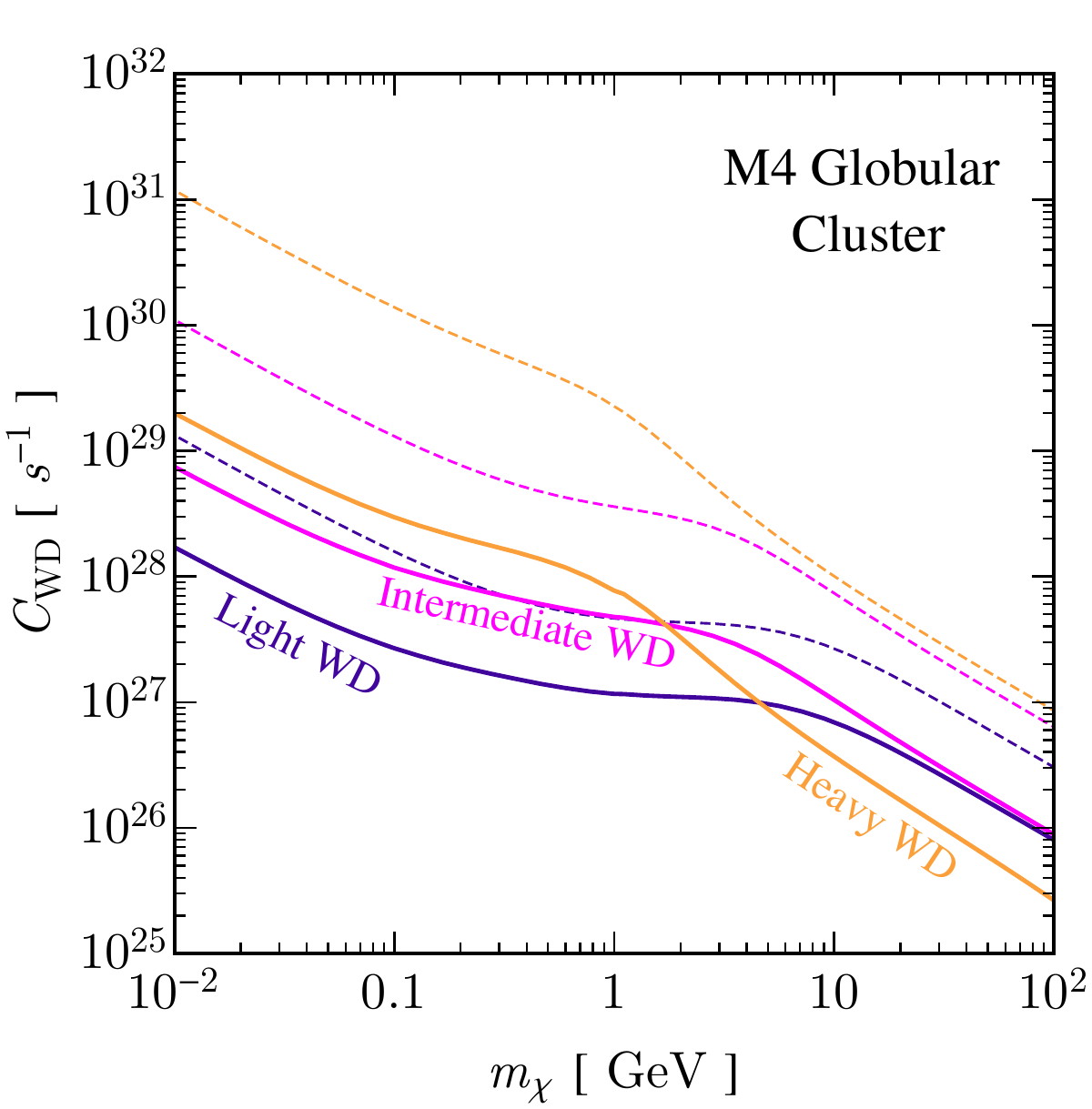}
    \caption{\textbf{Left:} Capture rates for our three benchmark white dwarfs at $1 \ \rm pc$ from the Galactic center, assuming an NFW profile and a DM-nucleon cross section $\sigma_{n\chi} = 10^{-45} \ \rm cm^2$. For each case, we show our capture rate which includes the white dwarf ions' zero-point energy (\textbf{solid}), as well as the previous treatment which assumes the ions are at rest (\textbf{dashed}). \textbf{Right:} Same as left panel, but for the M4 Globular Cluster under the common assumption of large DM content, see text for details.}
    \label{fig:capratecomp}
\end{figure}

Figure~\ref{fig:capratecomp} illustrates the difference between our improved treatment of DM capture in white dwarfs compared to previous works, in the context of two astrophysical systems: the inner Milky Way Galaxy (which we use in this work), and the globular cluster M4 (used dominantly in other works for white dwarf heating searches). In both cases, the capture rates are computed for all three benchmark white dwarfs in Tab.~\ref{tab:wds}, assuming a DM-nucleon cross section $\sigma_{n\chi} = 10^{-45} \ \rm cm^2$, and a temperature of $T_{\rm WD} = 10^5 \ \rm K$. For the inner Galaxy case, we chose a DM density $\rho_\chi \simeq 9322.7 \ \rm GeV / cm^{3}$ corresponding to an NFW profile at $r = 1 \ \rm pc$ from the Galactic center, and used the velocity parameters outlined in Sec.~\ref{sec:veldist}. For M4, we use the following parameters as per Ref.~\cite{Bell:2021fye}: $\rho_\chi = 798 \ \rm GeV / cm^{3}$, $v_{\rm WD} = 20 \ \rm km / s$ and $v_d = 8 \ \rm km / s$. Note that the density for M4 is subject to substantial systematics, and these are not robust values~\cite{Bell:2021fye,Moore:1995pb,Saitoh:2005tt}. We use these commonly assumed M4 values as an example globular cluster and to compare directly to their use in the literature; better clusters may be identified in the future with more precise DM content measurements.  

In Fig.~\ref{fig:capratecomp}, the dashed lines are the zero-temperature capture rates computed as per Ref.~\cite{Bell:2021fye}, which assumes the white dwarf ions to be at rest. This corresponds to replacing Eq.~\eqref{eq:cap2} with
\begin{equation}
    \Omega^-(w) = \frac{4 \mu_+^2}{\mu w} n_N(r) \int_{w \frac{|\mu_{-}|}{\mu_{+}}}^{v_{\rm esc}} dv \, v \, \frac{d\sigma_{N\chi}}{d\cos\theta}~,
    \label{eq:cap4}
\end{equation}
and taking $\langle v^2_N \rangle = 0$ in Eq.~\eqref{eq:relvdist}. However, as white dwarfs are extremely dense plasmas, the relevant energy scale at low temperatures is the ion plasma frequency, rather than the temperature of the white dwarf. Compared to taking only the ions at rest, our calculation, which is shown as the solid lines in Fig.~\ref{fig:capratecomp}, includes the quantum effects due to the dense white dwarf plasma. Including the zero-point motion is like considering a classical Maxwell-Boltzmann distribution for a substantially hotter white dwarf, despite the white dwarf actually being relatively cold;  see Appendix~\ref{app:RMSvel} for more details. 

In Fig.~\ref{fig:capratecomp}, compared to treating the ions as being at rest, we see there can be a substantial difference in the predicted DM capture rates. Firstly, in the left panel for the light white dwarf, we see that including the zero-point motion of the ions can increase the capture rate for the lightest DM masses plotted. This enhancement arises because the lower average DM kinetic energy leads to less suppression in the capture rate, and quantum effects increase the range of velocities contributing to capture ($cf.$ Fig.~\ref{fig:momdist}). For other white dwarf masses, this enhancement also can occur, however this happens at even lighter DM masses than shown. Secondly, as the DM mass increases, the enhancement disappears and the capture rate becomes suppressed relative to the treatment that assumes targets at rest. The reason for this decrease is due to the increasing DM kinetic energy leading to a regime where not all scatters result in a net energy loss for the DM. Thirdly, at high DM masses, there is a constant offset for all the white dwarf masses. The offset is constant at high DM masses, but not lower DM masses. This is because at lower DM masses other competing effects also affect the capture rate; namely for light DM masses the ion's kinetic energy dominates the center-of-mass energy. Still, in the heavy DM limit the solid line does not asymptote to the dashed line, because less collisions result in a net energy loss for the DM with the ion velocity included.

When $\langle v^2_N \rangle \gg v_d^2$, the gap between our treatment which includes the zero-point energy for the ions, and a capture treatment which assumes that the white dwarf ions are at rest, can be significant. This is why the difference between solid and dashed lines increases from the light to the heavy benchmark (higher density and therefore higher $\langle v^2_N \rangle$), as well as why it becomes more pronounced in the case of M4 (lower DM velocity dispersion compared to the Galactic center). In this last case, the difference can reach about two orders of magnitude for the heavy white dwarf benchmark.

While for lower masses we expect the capture rate to be enhanced by the effects we described above, it is also worth pointing out that we have not included any collective modes, which can become relevant in white dwarfs for DM masses below about 100 MeV depending on the DM particle model~\cite{DeRocco:2022rze}. For our DM constraints shown shortly, we do not enter the regime where collective effects are important, as we will focus on DM sensitivities above about 100 MeV.

\subsection{Maximum Capture Rate for a Single White Dwarf}

\begin{figure}[t!]
    \centering
    \includegraphics[width=\textwidth]{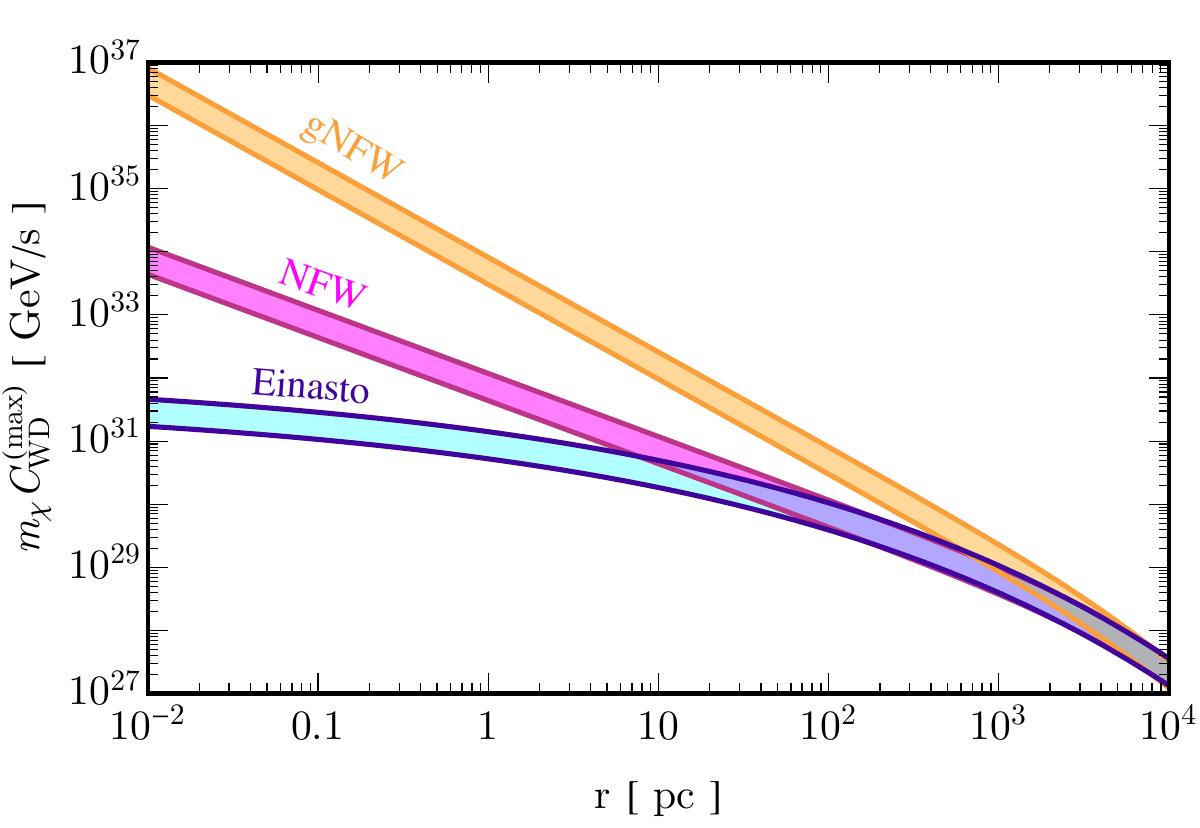}
    \caption{Maximum DM mass capture rate for white dwarfs as a function of the Galactocentric radius in the Milky Way Galaxy, for DM density profiles as labelled. The shaded bands cover the variation across the different white dwarf sizes in our benchmarks.}
    \label{fig:C_max}
\end{figure}

The maximum capture rate is obtained in the limit $\sigma_{n\chi} \rightarrow \infty$, where the white dwarf becomes optically thick and all incoming DM particles are captured near its surface. Coincidentally, quantum effects on the ion velocity distribution vanish near the surface because of the low density, recovering the expression \cite{Bell:2021fye}
\begin{equation}
    C_{\rm WD}^{\rm (max)} = \frac{\pi R_{\rm WD}^2 \rho_\chi}{3 v_{\rm WD} m_{\chi}} \left[ \left(3v_{\rm esc}^2(R_{\rm WD})+ 3 v_{\rm WD}^2 + v_d^2\right) {\rm erf}\left(\sqrt{\frac{3}{2}} \frac{v_{\rm WD}}{v_d}\right) + \sqrt{\frac{6}{\pi}} v_{\rm WD} v_d \exp\left(-\frac{3 v_{\rm WD}^2}{2 v_d^2}\right)\right].
    \label{eq:CWD_max}
\end{equation}
Figure~\ref{fig:C_max} shows the maximum capture rate for a single white dwarf as a function of the Galactic radius in the Milky Way Galaxy, for our three benchmark DM density profiles NFW, gNFW, and Einasto, as labelled. We show capture rates for our two extremal benchmark white dwarfs (see Sec.~\ref{sec:wdcomp} for modeling assumptions); the range between the extremal white dwarf benchmarks will approximately cover the expected variation in the Milky Way. The plot illustrates the expected variation in the signal size with DM density, which is not well known in the inner Galaxy. Our capture rates in this maximum scenario are very large, and will lead to substantial expected fluxes above the sensitivity of existing gamma-ray telescopes, as we will discuss shortly.

While the maximum capture rate is not technically reached until the cross section is infinite, we can approximate what cross section corresponds to a capture rate within a few tens of percent of the maximum.  We approximate the maximum capture cross section as the value for which $C_{\rm WD}$, as given in the optically thin limit by Eq.~\eqref{eq:cap1}, equals $C_{\rm WD}^{\rm (max)}$ for a given white dwarf benchmark. If the DM capture rate at this cross section is not sufficiently large to produce a detectable signal, increasing the cross section further will not aid detection, as the geometric capture rate of all DM passing through has already been approximately reached. Because our capture rates including the ion's zero-point energy are suppressed relative to the treatment assuming the ions are at rest, but the maximum capture rate is unchanged, our estimated maximum capture cross section for a given DM mass is higher than the values computed in Ref.~\cite{Bell:2021fye}, by the inverse of such suppression factor.

\section{Annihilation Flux and Gamma-Ray Detection}
\label{sec:detection}

\subsection{Dark Matter Thermalization and Annihilation Rate}
The total annihilation rate of captured DM within white dwarfs depends on the annihilation cross section, which determines the likelihood of two DM particles annihilating upon interaction. It also depends on their spatial distribution inside the star, which determines the frequency at which DM particles interact. The latter depends on the DM-nucleon cross-section, as this quantity governs how quickly the DM can lose energy and attain equilibrium with the object. 

In the weak interaction regime we consider, the thermalization of captured DM will occur in two stages. Initially, the DM particles will be bound to the white dwarf, with wide orbits that intersect the object but have turning points much larger than its radius. Due to the energy lost in each star crossing, the turning point of such orbits gradually contracts until the DM particles are fully contained within the white dwarf volume. Even for the large DM masses and small cross sections we probe, this first thermalization stage is completed on a short timescale. Following the estimates of Refs.~\cite{Kouvaris:2010jy,Acevedo:2019gre}, and choosing parameters $m_\chi \simeq 40 \ \rm TeV$ and DM-nucleon cross section $\sigma_{n\chi} \simeq 7 \times 10^{-47} \ \rm cm^2$ which correspond to the longest timescale in our detectable parameter space, we find the longest time for a captured DM particle to be fully trapped inside is about $3 \times 10^{-6}\ \rm Gyr$. Once this stage is complete, further loss of energy proceeds until the DM reaches a distribution with a virial radius that is much smaller than the white dwarf radius for the DM masses we consider. 

The timescale for capture and annihilation to reach equilibrium is
\begin{align}
    t_{\rm eq} = \sqrt{\frac{V}{\langle \sigma_{\rm ann} v_{\rm rel} \rangle \, C_{\rm WD}}}~,
    \label{eq:t_eq}
\end{align}
where $V$ is the volume in which annihilation proceeds and $\langle \sigma_{\rm ann} v_{\rm rel} \rangle$ is the thermal average of the product of annihilation cross section $\sigma_{\rm ann}$ and the relative velocity of the DM particles, $v_{\rm rel}$. We expand the annihilation cross section in partial waves $\ell$ as 
\begin{equation}
    \langle \sigma_{\rm ann} v_{\rm rel} \rangle = \langle \sigma_{\rm ann} v_{\rm rel} \rangle_0 \, \sum_{\ell=0}^{\infty} a_\ell \, v_{\rm rel}^{2\ell}\, .
\end{equation}
To individually analyze the contribution of each annihilation mode, we will set the coefficient $a_\ell = 1$ for each mode of interest while keeping the other modes equal to zero. Furthermore, we will also conservatively set the annihilation volume equal to the full volume of our light white dwarf ($cf.$ Table~\ref{tab:wds}), which is the largest in size amongst our benchmark white dwarfs. The relative velocity is estimated as the average speed over one radial orbit once the DM has been fully trapped by the white dwarf, and is of order $10^{-2} \, c$ for our light white dwarf. With these considerations, the timescales for capture-annihilation equilibrium are
\begin{equation}
    t_{\rm eq} \simeq 
        \begin{cases}
      \ 6 \times 10^{-3} \ {\rm Gyr}  & \ \text{$s$-wave} \\
      \\
      \ 0.6 \ {\rm Gyr} \left(\dfrac{10^{-2} \, c}{v_{\rm rel}}\right) & \ \text{$p$-wave} \\
    \end{cases}
    \label{eq:t_eq_modes}
\end{equation}
where in all cases we have normalized $\langle \sigma_{\rm ann} v_{\rm rel} \rangle_0 = 3 \times 10^{-26} \ \rm cm^3 / s$. The capture rate, on the other hand, has been conservatively set to the lowest value for which we would get an observable gamma-ray signal. This occurs also for a DM mass $m_\chi \simeq 40 \ \rm TeV$ and a DM-nucleon cross-section $\sigma_{n\chi} \simeq 9 \times 10^{-47} \ \rm cm^2$, for our benchmark light white dwarf situated at $r = 1.5 \ \rm pc$ assuming a gNFW profile (see below). For the other DM profiles, we have verified that they produce shorter equilibration timescales. This is because although the other profiles have lower DM densities, they correspond to larger cross sections. Therefore, comparing the smallest cross section we reach, which occurs for the gNFW DM density profile, is the most conservative check.

The capture-annihilation equilibrium timescales we find in Eq.~\eqref{eq:t_eq_modes} indicate that our setup has ample sensitivity to DM models with both $s$- and $p$-wave annihilation channels. In fact, even the $d$-wave channel, which is rarely testable in indirect detection searches, can reach equilibrium for some parameters (though this does not happen at the extremal conservative parameters). It is also worth reiterating that we have been highly conservative when making these estimates. In particular, we expect DM annihilation to proceed within a much smaller volume than the one we have used, leading to a much more efficient equilibration with DM capture. Therefore, we assume the annihilation rate to be
\begin{equation}
    \Gamma_{\rm ann} = \frac{C_{\rm WD}}{2}~,
    \label{eq:ann_rate}
\end{equation}
where the factor 2 reflects the fact that each annihilation event depletes two DM particles.

\subsection{Gamma-Ray Energy Flux}

The gamma-ray energy flux from a single white dwarf is calculated as \cite{Leane:2017vag}
\begin{equation}
\label{eq:flux}
    \left(E^2 \frac{d\Phi}{dE}\right)_{\rm WD} = \frac{\Gamma_{\rm{ann}}}{4\pi D^2}\, \times E^2 \frac{dN}{dE}\times {\rm{Br}}(\phi\rightarrow {\rm SM})\times P_{\rm{surv}} \, ,
\end{equation}
where we fix $D = 8 \ \rm kpc$ as the approximate distance between the white dwarfs in the central parsec of the Milky Way and the Earth, $E^2\frac{dN}{dE}$ is the gamma-ray spectrum per DM annihilation, and $\rm{Br}(\rm{\phi}\rightarrow \rm{SM})$ is the branching ratio of the mediator to SM particles. The probability of the mediator surviving to be detectable at Earth's position, $P_{\rm{surv}}$, is related to the decay length $L$ of the mediator by \cite{Leane:2017vag}
\begin{equation}
\label{eq:Psurv}
    P_{\rm{surv}} = \exp\left(-\frac{R_{\rm WD}}{L}\right) - \exp\left(-\frac{D}{L}\right), 
\end{equation}
where $R_{\rm WD}$ is the radius of the white dwarf. Eq.~\eqref{eq:Psurv} implies that $P_{\rm{surv}}$ is approximately one for $ R_{\rm WD} \lesssim  L \lesssim D$, which spans an extreme 14 orders of magnitude range of mediator decay lengths, so we take $P_{\rm{surv}}=1$. Other decay lengths may be probed, but will lead to decreased sensitivity and smaller values of $P_{\rm{surv}}$. We assume that the mediators escape the white dwarf without attenuation. This is a reasonable assumption for weakly coupled mediators, which is our scenario of interest.

To not specify a particle model, the branching ratio into photons is set to one, but other values can still provide strong sensitivity and can be simply rescaled accordingly. For DM annihilation into mediators that decay directly into photons, the spectrum is box shaped and given by \cite{Ibarra_2012}
\begin{equation}
    \frac{dN}{dE} = \frac{4}{\Delta E}\Theta(E-E_{-})\Theta(E_{+}-E)~,
\end{equation}
where $\Theta$ is the Heaviside function, $\Delta E = E_{+}-E_{-} = \sqrt{m_{\chi}^2-m_{\phi}^2}$ and 
\begin{equation}
    E_{\pm} = \frac{m_{\chi}}{2}\left(1 \pm \sqrt{1-\frac{m_{\phi}^2}{m_{\chi}^2}}\right),
\end{equation}
where $m_{\phi}$ is the mass of the mediator. While our search will yield results for a wide range of mediator masses, for demonstration purposes in our results below we will take $m_\phi = 10^{-3} \, m_\chi$. Any mediator mass leading to a decay length $R_{\rm WD} \lesssim L \lesssim D$ will also produce $P_{\rm{surv}}\simeq 1$ and so will provide essentially identical bounds. The lower end of the box spectra does change more notably with mediator mass, however it will still not affect the constraints if the mediator mass corresponds to $P_{\rm{surv}}\simeq 1$, as it is only the peak of the gamma-ray box spectra that sets the bound, which is approximately the same for all boosted mediators.

Following Ref.~\cite{Leane:2021ihh}, the resulting signal from the white dwarf population in the inner Galaxy is obtained from the volume integral
\begin{equation}
    \left(E^2 \frac{d\Phi}{dE}\right)_{\rm tot} = \int^{r_{\rm max}}_{r_{\rm min}} n_{\rm WD}(r) \, \left(E^2 \frac{d\Phi}{dE}\right)_{\rm WD} \, 4 \pi r^2 \, dr   ~, 
    \label{eq:tot_flux}
\end{equation}
where $n_{\rm WD}(r)$ is the number density of the white dwarf population at Galactic radius $r$ as per Eq.~\eqref{eq:nsc_pw}. We integrate from $r_{\rm min} = 10^{-2} \ \rm pc$ to $r_{\rm max} = 1.5 \ \rm pc$. The lower cutoff is chosen to avoid the region closest to Sag.~A$^*$, where the uncertainties in the DM and stellar velocity dispersions are particularly large. The upper cutoff corresponds to the maximum projected distance of 1.5 pc considered in Ref.~\cite{2023ApJ...944...79C}. We emphasize
that greater sensitivity can be obtained by extrapolating the power law to Galactocentric distances larger than 1.5 pc. In particular, our limits assuming a cored Einasto profile would
extend into the sub-GeV range in this case.

\subsection{Gamma-Ray Detection}

The Fermi Gamma-Ray Space Telescope (Fermi), and the High Energy Stereoscopic System (H.E.S.S.), have high-quality measurements of Galactic center gamma rays. Fermi, through its Large Area Telescope (LAT) instrument, has good sensitivity to gamma rays up to $\mathcal{O}$(100) GeV, and so will be relevant for annihilation of DM with masses up to this energy. H.E.S.S. is a system of imaging atmospheric Cherenkov telescopes and has sensitivity to gamma rays with energies of a few tens of GeV up to 100 TeV, and so will be used to obtain our TeV-scale DM sensitivity. While other gamma-ray telescopes operate in the TeV energy range, H.E.S.S. has the best Galactic center region exposure due to its southern hemisphere location. Other telescopes such as the High-Altitude Water Cherenkov Observatory (HAWC) and the Very Energetic Radiation Imaging Telescope Array System (VERITAS) lie in the northern hemisphere, and have poor exposures of the Galactic center region. Note that there are also complementary celestial-body searches using gamma rays from the Sun, and in that case TeV gamma rays require use of the HAWC Observatory, as atmospheric
Cherenkov telescopes (like H.E.S.S. and VERITAS) would be \textit{fried} if pointed at the Sun. Fermi can be used for both the Sun and the Galactic center region, as it has all-sky exposure, and detects gamma rays through pair conversion.

Throughout the DM mass range we explore, the flux calculated from Eq.~(\ref{eq:tot_flux}) is compared to the gamma-ray flux data from Fermi and H.E.S.S observations \cite{Malyshev:2015hqa}. We then exclude DM-nucleon cross sections for which the resulting flux would be greater than the observed flux in any energy bin, which is a very conservative limit. Modeling the astrophysical backgrounds would provide stronger DM limits than what we present here.

\section{Dark Matter Parameter Space Constraints}
\label{sec:DMconstraints}
\subsection{Results}

\begin{figure*}[t!]
    \centering
    \includegraphics[width=\textwidth]{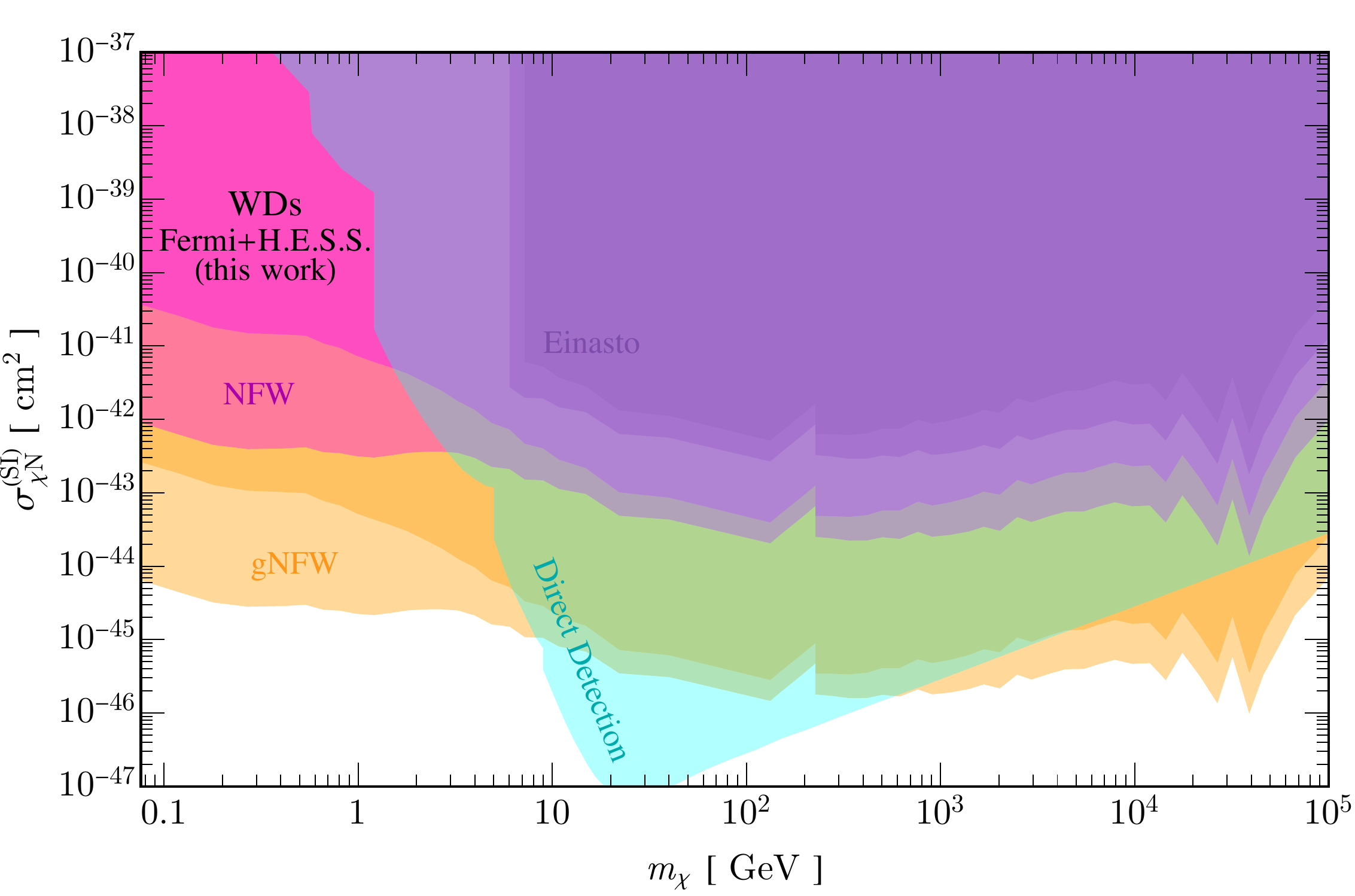}
    \caption{Spin-independent DM-nucleon cross-section limits derived from Fermi-LAT and H.E.S.S. gamma-ray data using white dwarfs in the inner Milky Way, for NFW, gNFW and Einasto DM profiles as labelled. The darker and lighter shaded regions for a given DM profile correspond to the conservative or optimistic result respectively across our white dwarf benchmarks. Complementary constraints from direct detection are also shown, see text for details.} 
    \label{fig:crossseclimits}
\end{figure*}

Figure~\ref{fig:crossseclimits} shows our calculated spin-independent cross-section sensitivity as a function of DM mass, for three Galactic DM distributions (gNFW, NFW, and Einasto). The bands for each case illustrate the variation of these limits with the white dwarf's physical properties, and are detailed further below along with other uncertainties in the white dwarf modeling. We also show current leading limits from direct detection experiments, which are LZ \cite{LZ:2022ufs}, PandaX-4T \cite{PandaX-4T:2021bab}, DarkSide-50 \cite{DarkSide-50:2022qzh}, and a low-mass search in Xenon-1T accounting for the Migdal effect \cite{XENON:2019zpr}. Depending on the DM density profile, our white dwarf bounds can outperform these searches below 10 GeV masses as well as above a few hundred GeV. The improvement in sensitivity is especially stark in the sub-GeV region, where direct detection sensitivity is suppressed due to limited recoil thresholds.

In Fig.~\ref{fig:crossseclimits}, the shape of the cross section limits is driven by both the capture rate scaling with DM mass and the Galactic center gamma-ray data from Fermi and H.E.S.S.. At around 230 GeV DM mass, the kink in the curve is due to the H.E.S.S. sensitivity taking over from the Fermi sensitivity. The dips near a few tens of TeV DM mass are due to the H.E.S.S. measured background gamma-ray flux being particularly low at these corresponding energies, leading to the strongest bounds. Approaching 100 TeV DM mass, the bounds weaken rapidly due to the peak of the gamma-ray spectrum moving out of the H.E.S.S. energy detection range. The edge of the gamma ray spectrum is still detectable, but leads to smaller fluxes and therefore weaker constraints. Above about 10 TeV DM mass, multiscatter capture also becomes important. However, we have conservatively only considered a single scatter formalism even in this mass range, and so the limits rise faster than they would in a more complete treatment.

Other than white dwarfs, another celestial body population that is able to probe sub-GeV DM through gamma-rays are brown dwarfs \cite{Leane:2021ihh}. Our sub-GeV limits are stronger than those derived from brown dwarfs in Ref.~\cite{Leane:2021ihh} by a few orders of magnitude in the case of spin-independent scattering. These improvements occur due to the higher number density of white dwarfs at the Galactic center, and the higher white dwarf capture efficiency at lower cross sections. Note that Ref.~\cite{Leane:2021ihh} focuses on the spin-dependent scattering parameter space. While our white dwarf limits are stronger than the spin-independent limits expected for brown dwarfs, in the case of spin-dependent scattering the brown dwarf search in Ref.~\cite{Leane:2021ihh} is superior. This is because spin-dependent scattering only occurs with elements which have non-zero nuclear spin. White dwarfs contain dominantly carbon and oxygen, which have zero nuclear spin, leading to a negligible spin-dependent scattering rate. Therefore, whether the white dwarf or brown dwarf population gamma-ray search provides the stronger limits will depend on the particle physics model in question, that is whether DM exhibits spin-independent or spin-dependent scattering. The Galactic center neutron star population was also proposed in Ref.~\cite{Leane:2021ihh} as DM gamma-ray emitters. However, as neutron stars are much smaller, their maximal capture rate is smaller, and they therefore only produce a sufficiently large signal in a limited part of the DM parameter space~\cite{Leane:2021ihh}.

Limits on the DM-nucleon scattering parameter space have also been calculated using gamma rays from the Sun~\cite{Leane:2017vag,Arina:2017sng,Albert:2018jwh,Bell:2021pyy} and Jupiter~\cite{Leane:2021tjj}, and depending on the DM model may be competitive with our limits in Fig.~\ref{fig:crossseclimits}. The main benefit in using a Galactic center population search (be it with brown dwarfs, white dwarfs, or neutron stars) compared to a local position search is the larger range of mediator decay lengths that can be detected. For many of the extreme decay lengths which can be probed with our search with strong sensitivity, the corresponding Sun or Jupiter bounds do not exist. We therefore do not show these limits in Fig.~\ref{fig:crossseclimits}, as they are not directly comparable, but they should be kept in mind as an alternative search strategy. 

Compared to existing searches, our white dwarf search has other additional advantages. These include the fact that capture-annihilation equilibrium is possible for very small cross sections, even for highly suppressed annihilation modes such as $p$-wave or $d$-wave annihilation. Another benefit of using white dwarfs over the Sun or Jupiter is the low white dwarf evaporation mass. However, while white dwarfs can access much lower DM masses, the exact evaporation mass is highly model dependent. For example, attractive long-range forces can substantially decrease the DM evaporation mass~\cite{Acevedo:2023owd}, making Jupiter and the Sun potentially competitive at even sub-MeV masses, depending on the DM model parameters. Local position searches are also not subject to the same astrophysical uncertainties as a Galactic center population search, however conservative choices can be made to still reveal substantial complementary sensitivity.

It is worth reiterating that here we have assumed that the dark mediator decays directly into gamma rays. While this is the most optimistic scenario, it is not the only possible one. Other final states with varied branching ratios generally decrease the sensitivity. However, given the extreme increase in cross section sensitivity we find compared to existing searches, less optimistic scenarios will also be detectable with our search. Indeed, while we have not specified a specific particle model for our bounds, there are a wide range that can be applicable to this white dwarf scenario. One example is a scalar-pseudoscalar model, similar to the one considered for the Sun in Ref.~\cite{Arina:2017sng}. Several other applicable model examples are also considered in the context of the Sun in Ref.~\cite{Batell:2009zp}.

\subsection{Discussion of Astrophysical Uncertainties}

\subsubsection{Dark Matter Uncertainties}

In Fig.~\ref{fig:crossseclimits}, our DM density profiles cover a range of what is considered in the literature, and illustrate that the DM density is a substantial systematic for this inner Galaxy search (as per effectively all Milky Way indirect DM detection searches). In any case, even taking up to about 6 orders of magnitude uncertainty in the DM density in the inner Galaxy (see Fig.~\ref{fig:C_max}), across cored and cuspy profiles, we show that DM can still be probed with this search. As expected, the cuspier profiles (gNFW and NFW) provide stronger limits than the cored profile (Einasto), due to the greater DM density at the Galactic center predicted by the cuspier profiles. The most conservative case is therefore given by the Einasto profile. The bounds for this profile are shown only for DM masses above about 6 GeV, because these are the only DM masses for which any cross section corresponds to a capture rate large enough to be detected. However, sub-GeV sensitivity could be attained with an Einasto DM profile by taking a less conservative approach than we do here, and integrating the white dwarf density out to larger Galactic radii than 1.5 pc. 

The most optimistic case is provided by the gNFW profile with $\gamma=1.5$, and such cuspy distributions are well motivated as they are commonly predicted when including baryonic effects, $i.e.$ adiabatic contraction. In this case, our sub-GeV limits are stronger than leading direct detection experiments by at least $\sim 4-9$ orders of magnitude. Even in a more conservative case, considering the NFW profile with $\gamma=1$, our sub-GeV bounds are stronger than direct detection by $\sim 2-7$ orders of magnitude. Note that while we have shown a range of DM density profile benchmarks, these do not necessarily cover all possibilities. It is plausible that the DM density profile is somewhat different to the cases we show here, as it is still not robustly known or measured. However, the cases we show do cover a substantial variation in the DM density profiles considered in the literature, and do still produce strong sensitivity.

\subsubsection{White Dwarf Uncertainties}
The sensitivity bands for a given DM profile in Fig.~\ref{fig:crossseclimits} show the range of our results spanned by the three benchmark white dwarfs we consider. These are constructed by taking the minimum and maximum from the cross-section limits generated by each individual benchmark, where each extreme case was obtained assuming all Galactic white dwarfs have the same physical properties ($i.e.$ mass, radius, central density). Our light white dwarf benchmark drives the upper limit of the band (the most conservative case) from the sub-GeV regime up until $m_\chi \simeq 230 \ \rm GeV$. Past this value, the most conservative limits are actually drawn from our heavy benchmark. This is because heavier white dwarfs have denser cores with larger plasma frequencies, leading to a suppression such that the heavy benchmark is more inefficient at capturing DM in this high mass end despite having more target nuclei. For the lower limit of the band (the most optimistic case), constraints in the sub-GeV regime are driven by our heavy benchmark white dwarf for masses up to $m_\chi \simeq 3 \ \rm GeV$. Beyond this value, the lower limit is determined by our intermediate benchmark, as this white dwarf configuration is overall more effective at capturing dark matter than both the light and heavy cases ($cf.~$ Fig.~\ref{fig:capratecomp}). This is due to the increased number of targets compared to the light benchmark, combined with having a milder suppression compared to the heavy benchmark. More detailed modeling and observations of the white dwarf mass distribution in the Galactic center will lead to a reduction in the band width. 

We also comment on our assumptions about the white dwarf number density distribution in the central parsec of the Milky Way, $cf.$ Eq.~\eqref{eq:nsc_pw}. Reported values for the power law index $\alpha$ range from about $1.0$ \cite{Panamarev:2018bwq}, to about $1.4$ \cite{Alexander:2008tq}. Within this range, our results only vary from a few percent in the case of our Einasto profile, to about sixty percent for the gNFW profile. On the other hand, to fix the normalization of Eq.~\eqref{eq:nsc_pw}, we used state-of-the-art predictions for the white dwarf abundance from Ref.~\cite{2023ApJ...944...79C}, which incorporated recent observations of super-solar metallicity in this region. It is worth pointing out that this estimate depends on the initial-to-final mass relation utilized to model the star formation history. At present, only one is available for high metallicity stars \cite{2015MNRAS.451.4086S}. Advancements in the modeling of star formation and evolution in this high metallicity regime will lead to more accurate determinations of the number of stellar remnants in the future.

Finally, we discuss the impact of dark matter kinetic heating on our estimates. Depending on the cross section and DM halo profile, we find white dwarfs in the Milky Way's inner parsec can be kinetically heated up to $T_{\rm WD} \simeq 10^7 \ \rm K$. We verified that this temperature increase does not lead to more than a few percent variation in our capture rate and thus the cross-section limits we show in Fig.~\ref{fig:crossseclimits}. We also note that this temperature increase is not significant enough for corrections to the equation of state to become relevant for the range of white dwarf masses and composition we consider~\cite{deCarvalho:2013rea}.

\section{Summary and Conclusions}
\label{sec:conclusion}

White dwarfs are excellent DM detectors. They are incredibly dense, leading to efficient capture for small cross sections, while also having large surface areas, leading to large capture rates. Previous studies of DM in white dwarfs have largely focused on DM heating of white dwarfs in globular clusters such as Messier 4, where the DM content is not well understood. We instead targeted the Milky Way Galactic center white dwarf population, where even for a range of DM density distributions, a substantial DM signal could arise. We explored the scenario where DM is captured and annihilates into boosted or long-lived mediators, which can escape the Galactic white dwarf population and decay into detectable gamma rays. We used Fermi and H.E.S.S. Galactic center gamma-ray data to place strong limits on the resulting gamma-ray signal, and constrained the DM-SM scattering cross section for DM masses as low as 80 MeV. We have shown that for a cuspy halo profile, spin-independent cross section limits from DM capture by white dwarfs can outperform direct detection limits by up to nine orders of magnitude for sub-GeV DM masses. Above a few hundred GeV, our white dwarf limits for that same halo profile are also stronger than world-leading direct detection limits.

We also presented an improved treatment of the DM capture rate in white dwarfs, which accounts for quantum effects of the ions when the white dwarf temperature falls below the ion plasma frequency. The zero-point motion of the ions in this regime results in a significantly higher velocity compared to the usual thermal estimate used in previous works. We then showed that this effect produces lower capture rates in the Galactic center by a factor of up to about five in our DM mass range, depending on the white dwarf mass. While the focus of our work was on DM in Galactic center white dwarfs, DM in globular cluster white dwarfs have been often considered in the literature. The capture framework we discussed in this work is also applicable to those results -- for globular clusters, the difference between our treatment and neglecting the ion thermal velocity is even larger, due to the DM velocity dispersion often being very low in globular clusters, and therefore the ion velocity being potentially larger than the DM velocity. For the example case of the commonly considered M4 globular cluster, we found that our improved ion velocity treatment leads to DM capture rates up to two orders of magnitude smaller than was previously considered.

In the future, sensitivities for DM searches in white dwarfs will be improved with advances in our understanding of the DM content both in our galaxy and in globular clusters. Our strongest constraints assume that the center of the Milky Way is very DM rich, but the systematic uncertainty is high given the extensive range of possible DM density profiles. Although our sensitivity is strong for a range of DM profiles, an improved understanding of the DM density in our galaxy could reduce the uncertainty in our cross section limits by several orders of magnitude. It is also possible that better understanding the DM content in globular clusters could lead to improved sensitivity of both white dwarf heating and gamma-ray signals. In particular, a positive DM detection is possible if new clusters with high DM densities are discovered nearby, and any anomalous signals from gamma rays or heating were found. Another possibility for future work is to take advantage of nearby individual white dwarfs, which could lead to strong DM sensitivity. 

\section*{Acknowledgments} 

We thank Nicole Bell, Joseph Bramante, Giorgio Busoni, Christopher Cappiello, Zhuo Chen, Bernhard Mistlberger, Juri Smirnov, and Aaron Vincent for helpful discussions. JFA, RKL, and LSO are supported in part by the U.S. Department of Energy under Contract DE-AC02-76SF00515. LSO is supported by a Stanford Physics Department Fellowship.

\appendix

\section{Velocity Distribution for Ions in Cold White Dwarfs}
\label{app:RMSvel}
We derive the momentum distribution of the white dwarf ions at finite temperature, Eq.~\eqref{eq:momdist}, starting from the expression (see $e.g.$ Ref.~\cite{pathria2011statistical})
\begin{equation}
    \mathcal{F}_{\rm ion}(v_N) = \mathcal{Z}^{-1} \sum_n |\langle \mathbf{v}_N| n \rangle|^2\exp\left(-E_n/T_{\rm WD} \right)\,,
    \label{eq:app-veldistgen}
\end{equation}
where $T_{\rm WD}$ is the white dwarf temperature, and the sum is over all energy eigenstates $E_n$. We approximate the latter by those of a harmonic oscillator potential with a natural frequency equal to the ion plasma frequency $\omega_p$ of the medium, whereby $|n \rangle = |n_x,n_y,n_z\rangle$. This approximation is valid for white dwarfs sufficiently cold that their temperature is below $\omega_p$. In this thermodynamic state, the velocity dispersion of nuclei is determined by the energy scale $\omega_p$ rather than the temperature, as we show shortly. Note that $\mathcal{F}_{\rm ion}(v_N)$ will be only a function of $v_N = |\mathbf{v}_N|$. The partition function $\mathcal{Z}$ is
\begin{equation}
    \mathcal{Z} = \sum_n \exp\left(-E_n/T_{\rm WD} \right) = \frac{\exp\left(-3 \omega_p/2 T_{\rm WD}\right)}{\left(1-\exp(-\omega_p/T_{\rm WD})\right)^3}~.
\end{equation}
The squared matrix elements read 
\begin{equation}
    |\langle \mathbf{v}_N| n \rangle|^2 = \left(\frac{m_N}{\pi \omega_p}\right)^{3/2} \prod_{j = x,y,z} \frac{1}{2^{n_j} n_j!} \, H^2_{n_j} \! \left(\sqrt{\frac{ m_N}{\omega_p}}v_j \right) \, \exp\left(-\frac{m_N}{\omega_p} v^2_j\right)\,,
\end{equation}
where the functions $H_n(x)$ are Hermite polynomials of order $n$. The sum in Eq.~\eqref{eq:app-veldistgen} can be computed exactly by making use of the following completeness relation for Hermite polynomials (see p.~194 of Ref.~\cite{bateman}) 
\begin{equation}
    \sum_{n=0}^{\infty} \frac{H_n(x) H_n(y)}{2^n \, n!} z^n =\left(1-z^2\right)^{-1/2}\exp\left(\frac{2z}{1+z} xy - \frac{z^2}{1-z^2} (x-y)^2\right)~,
\end{equation}
with the identifications $z = \exp(-\omega_p/T_{\rm WD})$ and $x = y = \sqrt{m_N/\omega_p} \, v_j$ for each separate component $j$. Factorizing the sum into each cartesian component and using the above identity we obtain the distribution
\begin{equation}
     \mathcal{F}_{\rm ion}(v_N)= \left(\frac{m_N}{\pi \omega_p \coth{\left(\omega_p /2 T_{\rm WD}\right)}}\right)^{3/2} \exp\left[-\frac{m_N v_N^2}{\omega_p \coth\left(\omega_p/2T_{\rm WD}\right)}\right]~,
     \label{eq:momdist}
\end{equation}
with a mean squared velocity
\begin{equation}
    \langle v_N^2 \rangle = 4 \pi \int_0^\infty dv_N \, v_N^4 \, \mathcal{F}_{\rm ion}(v_N) = \left(\frac{3\omega_p}{2m_N}\right) \coth \left(\frac{\omega_p}{2 T_{\rm WD}}\right)~.
\end{equation}

\begin{figure}
    \centering
    \includegraphics[width=0.495\textwidth]{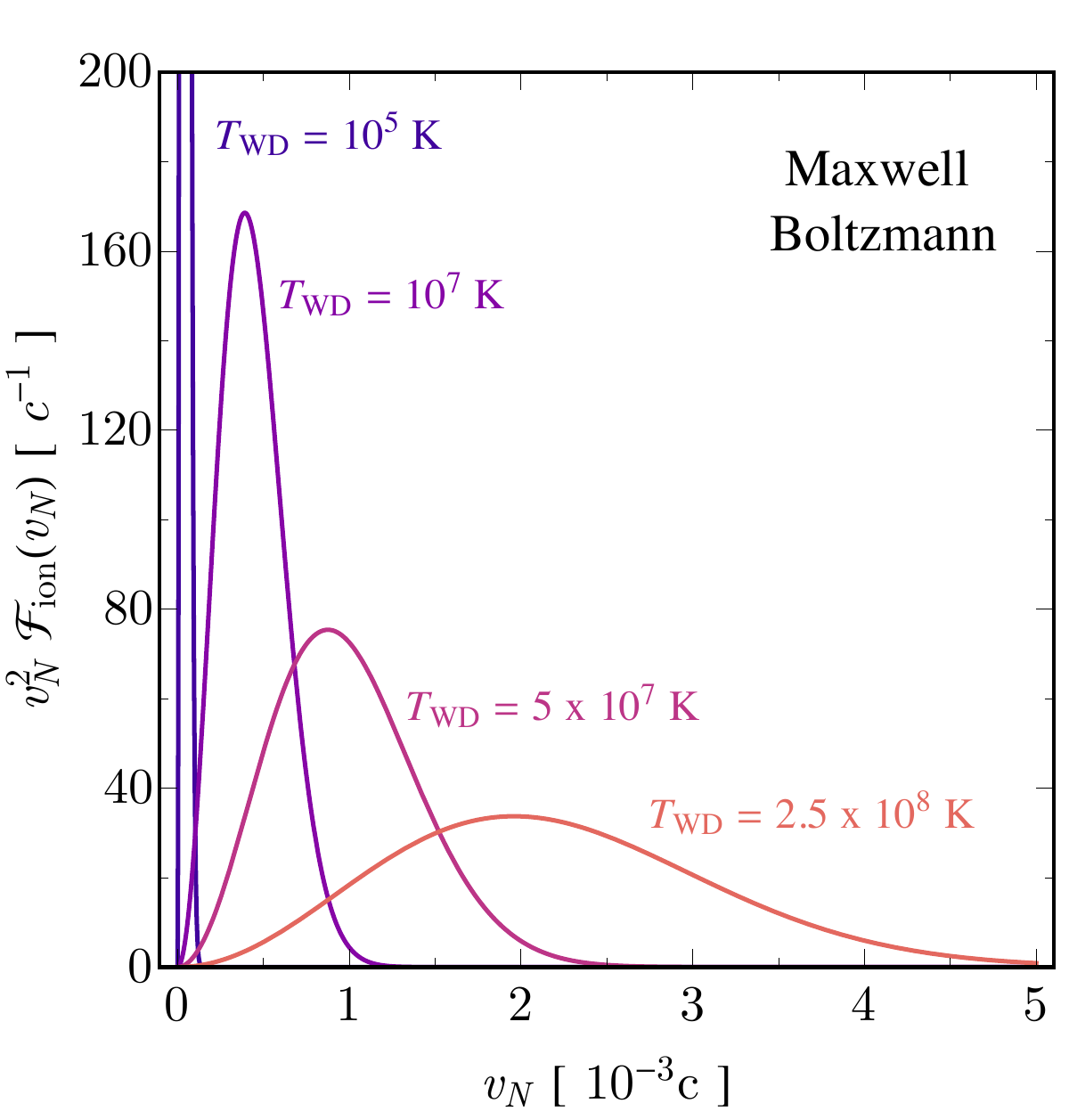}
    \includegraphics[width=0.495\textwidth]{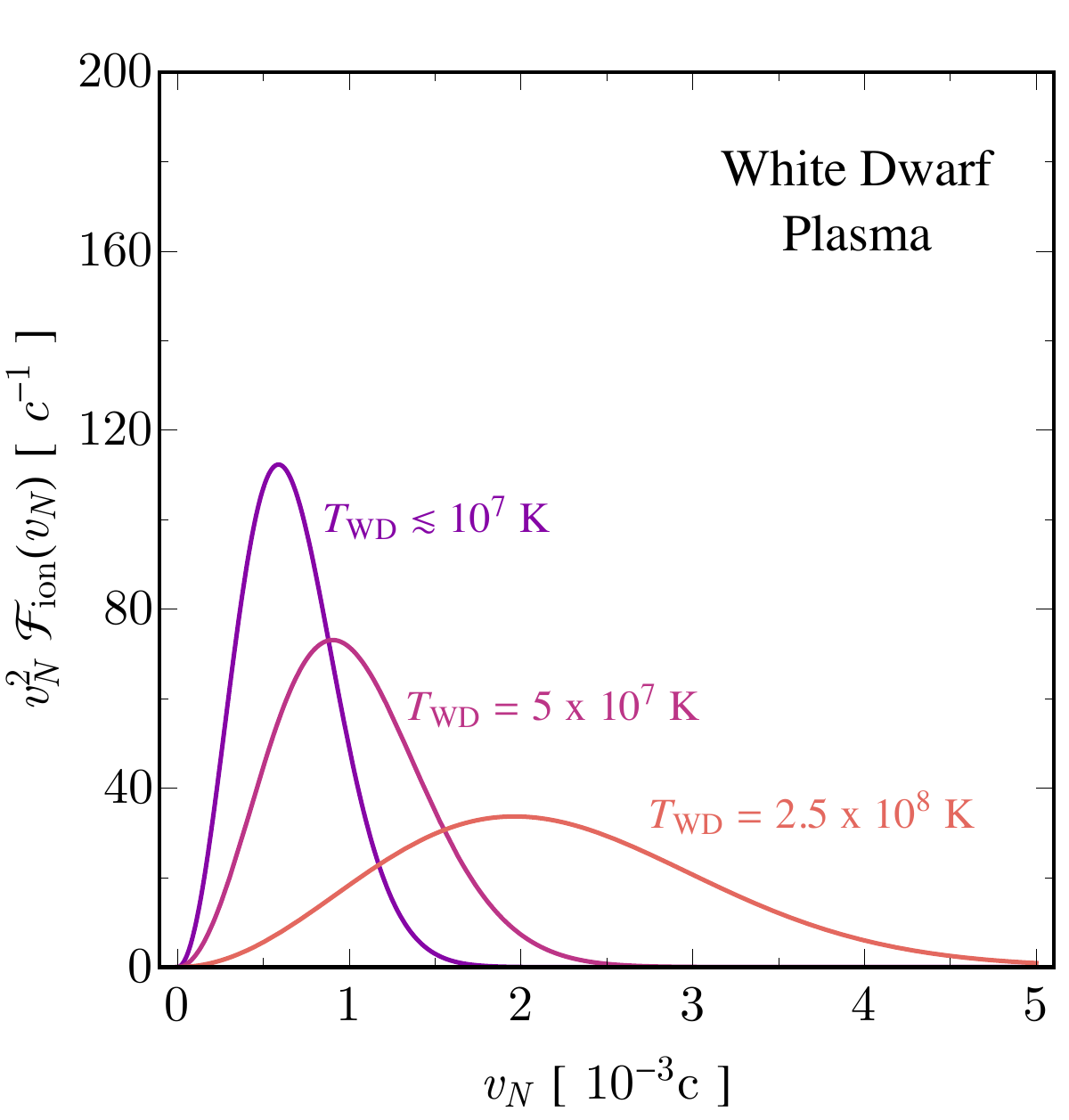}
    \caption{\textbf{Left:} Speed probability density of the white dwarf ions in the stellar rest frame using the classical Maxwell-Boltzmann distribution, assuming a pure carbon composition for temperatures as specified. This is the classical picture. \textbf{Right:} Same as left panel, but including the ion's zero-point motion (as per Eq.~\eqref{eq:momdist}) for an ion density of order $10^7 \ \rm g / cm^{3}$. This is the quantum picture.}
    \label{fig:app-momdist}
\end{figure}

Note that the ion plasma frequency depends on the local ion density, $cf.$ Eq.~\eqref{eq:ion_plasma_freq}. Thus, for old white dwarfs in the inner Galaxy, most of their volume will satisfy $T_{\rm WD} \ll \omega_p$ except very near the surface. In the regions where the plasma frequency dominates over temperature, only the ground state contributes to the sum in Eq.~\eqref{eq:app-veldistgen}, yielding
\begin{equation}
    \mathcal{F}_{\rm ion}(v_N) \xrightarrow[T_{\rm WD} \, \ll \, \omega_p]{} \left(\frac{m_N}{ \pi \omega_p}\right)^{3/2}\exp\left(-\frac{m_N v_N^2}{\omega_p}\right)~,
\end{equation}
with a mean squared velocity
\begin{equation}
    \langle v_N^2 \rangle \xrightarrow[T_{\rm WD} \, \ll \, \omega_p]{} \frac{3 \omega_p}{2m_N}~.
    \label{eq:v_disp_quant}
\end{equation}
Thus, the ion velocity distribution for old white dwarfs can be thought of being like a Maxwell-Boltzmann, but the plasma frequency replaces the temperature as the energy scale. 

As a final consistency check, note that in the limit that quantum effects are irrelevant, which will occur near the surface of the white dwarf, the above distribution asymptotes to a classical Maxwell-Boltzmann distribution
\begin{equation}
    \mathcal{F}_{\rm ion}(v_N) \xrightarrow[T_{\rm WD} \, \gg \, \omega_p]{} \left(\frac{m_N}{2\pi T_{\rm WD}}\right)^{3/2} \exp\left(-\frac{m_N v_N^2}{2 T_{\rm WD}}\right)~,
    \label{eq:app_MB_dist}
\end{equation}
with the mean squared velocity matching that of Ref.~\cite{Bell:2021fye},
\begin{equation}
    \langle v_N^2 \rangle \xrightarrow[T_{\rm WD} \, \gg \, \omega_p]{} \frac{3 T_{\rm WD}}{m_N}~.
  \label{eq:v_disp_class}
\end{equation}

Figure~\ref{fig:app-momdist} illustrates how our distribution, Eq.~\eqref{eq:momdist}, converges to the classical Maxwell-Boltzmann distribution given by Eq.~\eqref{eq:app_MB_dist} as the temperature increases, for a fiducial density $\rho_{\rm WD} = 10^7 \ \rm g / cm^{3}$. In consistency with the above, at high temperatures both distributions align closely. However, as the temperature decreases, Eq.~\eqref{eq:momdist-main} exhibits a significantly broader spread compared to the classical Maxwell-Boltzmann distribution. This is because the plasma frequency serves as a lower bound on the energy scale of the white dwarf ions. 

\bibliographystyle{JHEP}
\bibliography{dwarfs}	
\end{document}